\documentclass[10pt]{wlscirep}
\title{Weighted Uncertainty Relations}

\author[1, 2]{Yunlong Xiao}
\author[1, 3, *]{Naihuan Jing}
\author[2]{Xianqing Li-Jost}
\author[2, 4]{Shao-Ming Fei}
\affil[1]{School of Mathematics, South China University of Technology, Guangzhou 510640, China}
\affil[2]{Max Planck Institute for Mathematics in the Sciences, Leipzig 04103, Germany}
\affil[3]{Department of Mathematics, North Carolina State University, Raleigh, NC 27695, USA}
\affil[4]{School of Mathematical Sciences, Capital Normal University, Beijing 100048, China}

\affil[*]{Corresponding author: jing@ncsu.edu}

\begin{abstract}
Recently, Maccone and Pati 
have given two stronger uncertainty relations based on the sum of variances and one of them is nontrivial when the quantum state is not an eigenstate of the sum of the observables. We derive a family of
weighted uncertainty relations to provide
an optimal lower bound for all situations and
remove the restriction on the quantum state. Generalization to multi-observable cases is also given and an optimal lower bound for the weighted
sum of the variances is obtained in general quantum situation.
\end{abstract}
\begin{document}

\flushbottom
\maketitle

\thispagestyle{empty}

\section*{Introduction}

In Kennard's  formulation \cite{Kennard} of Heisenberg's uncertainty principle \cite{Heisenberg},
for any single quantum particle, the product of the uncertainties of the position and momentum measurements is at least half of the Planck constant
(see also the work of Weyl \cite{Weyl})
\begin{equation}
\Delta X\cdot\Delta P\geqslant \frac{\hbar}{2}.
\end{equation}

Later Robertson \cite{Robertson} derived the uncertainty principle for any pair of observables $A$ and $B$ with bounded spectrums:
\begin{equation}\label{e:Robertson}
\Delta A\cdot\Delta B\geqslant\frac{1}{2}|\langle [A, B]\rangle|,
\end{equation}
where $\Delta A^2=\langle A^2\rangle-\langle A\rangle^2$ is the variance 
of operator $A$ over the state $|\psi\rangle$.
Eq. (\ref{e:Robertson}) can be derived from a slightly strengthened inequality, the Schr\"{o}dinger uncertainty relation \cite{Schrodinger}
\begin{equation}
\Delta A^{2}\cdot\Delta B^{2}\geqslant|\frac{1}{2}\langle[A, B]\rangle|^{2}+|\frac{1}{2}\langle\{\widehat{A}, \widehat{B}\}\rangle|^{2},
\end{equation}
where
 $\widehat{A}={A}-\langle A\rangle I$ and $I$ is the identity operator.

All these inequalities \cite{Busch, Lahti} can be trivial even if $A$ and $B$ are incompatible on the state of the system $|\psi\rangle$, for instance, when $|\psi\rangle$ is an eigenstate of either $A$ or $B$. Despite of this, the variance-based uncertainty relations possess a clear physical meaning and have variety of applications in the theory of quantum information processing such as entanglement detection \cite{Hofmann, Guhne}, quantum spin squeezing \cite{Walls, Wodkiewicz, Wineland, Kitagawa, Ma}, and quantum metrology \cite{Giovannetti, Lloyd, Maccone}.

Recently Maccone and Pati have presented two stronger uncertainty
relations \cite{Pati} based on the sum of variances and their inequalities are guaranteed to be nontrivial when $|\psi\rangle$ is not a common eigenstate of $A$ and $B$. Though there are many formulations of the uncertainty relation in terms of the sum of entropic quantities \cite{Wehner, Coles}, Maccone and Pati's relations capture the notion of incompatibility except when the state is an eigenstate of the sum of the operators.
Their first relation for the sum of the variances is
\begin{equation}\label{e:MP1}
\Delta A^{2}+\Delta B^{2}\geqslant\pm i\langle[A, B]\rangle+|\langle\psi|A\pm iB|\psi^{\perp}\rangle|^{2}:=\mathcal{L}_{MP1},
\end{equation}
which is valid for any state $|\psi^{\perp}\rangle$ orthogonal to the state of the system $|\psi\rangle$ while the sign should be chosen so that $\pm i\langle[A, B]\rangle$ is positive. Denote the right-hand (RHS) of Eq. (\ref{e:MP1}) by $\mathcal{L}_{MP1}$. Their second
uncertainty relation also provides a nontrivial bound even if $|\psi\rangle$ is an eigenstate of $A$ or $B$:
\begin{equation}\label{e:MP2}
\Delta A^{2}+\Delta B^{2}\geqslant\frac{1}{2}|\langle\psi^{\perp}_{A+B}|A+B|\psi\rangle|^{2}:=\mathcal{L}_{MP2},
\end{equation}
where $|\psi^{\perp}_{A+B}\rangle\propto(A+B-\langle A+B \rangle)|\psi\rangle)$ is a state orthogonal to $|\psi\rangle$. It is easy to see that the RHS $\mathcal{L}_{MP2}$ of Eq. (\ref{e:MP2}) is nontrivial unless $|\psi\rangle$ is an eigenstate of $A+B$. Moreover, based on the same techniques,
Maccone and Pati also obtained an amended Heisenberg-Robertson inequality:
\begin{equation}
\Delta A\Delta B\geqslant\pm\frac{i}{2}\langle[A, B]\rangle/(1-\frac{1}{2}\mid\langle\psi|\frac{A}{\Delta A}\pm i\frac{B}{\Delta B}|\psi^{\perp}\rangle\mid^{2}),
\end{equation}
which reduces to Heisenberg-Robertson's uncertainty relation when minimizing the lower bound over $|\psi^{\perp}\rangle$, and the equality holds at the maximum. The goal of this paper is
to give a new method of measuring the uncertainties to remove the restriction on the bounds such as $\mathcal{L}_{MP2}$.

Actually, both the entropic uncertainty relations and the sum form of variance based uncertainty do not suffer from trivial bounds. Generalizing Deutsch's entropic uncertainty relation \cite{Deutsch},
Maassen and Uffink \cite{Maassen} used certain weighted entropic uncertainties to derive a tighter bound.
Adopting a similar idea to the uncertainty relations based on R\'{e}nyi entropy, we propose a {\it deformed uncertainty relation} to resolve the restriction of Maccone-Pati's variance based uncertainty relation. i.e. the new uncertainty relation will provide a nontrivial bound even when
the state is an eigenvector of $A+B$.
Moreover, we show that the original Maccone-Pati's bound is a singular case in our general uncertainty relation
and the usual sum of variances can be extracted from weighted sum of uncertainties.
Our work indicates that it seems
unreasonable to assume a priori that observables $A$ and $B$ have equal contribution to
the variance-based sum uncertainty relation.
Our family of uncertainty relations are proved to possess
an optimal bound in various situations according to the state of the system.
In particular, all previous important variance-based sum uncertainty relations are special cases
of our weighted uncertainty relation.

We remark that there is another approach of {\it measurement uncertainty} \cite{BLW1, BLW2} to the uncertainty principle
which deals with joint measurability and measurement-disturbance.
Our methods can also be used to generalize the joint measurability, also known as
{\it preparation uncertainty} \cite{BLW1}, and to obtain a tighter bound.

\section*{Results}

We first consider the weighted uncertainty relations based on the sum of variances of two observables, then generalize it into multi-observable cases.
All observables considered in the paper will be assumed to be non-degenerate on a finite-dimensional Hilbert space.
We will show that our weighted uncertainty relations give optimal lower bounds and all previous important variance-based sum uncertainty relations are special cases of the new weighted uncertainty relation.

\vspace{2ex}
\leftline{\textbf{Theorem 1} {\it For arbitrary observables $A$, $B$ and any positive number
$\lambda$, we have the following weighted uncertainty relation:}}
\begin{align}\label{e:ineq}
(1+\lambda)\Delta A^{2}+(1+\lambda^{-1})\Delta B^{2}
\geqslant&-2i\langle[A, B]\rangle+|\langle\psi|(A-iB)|\psi^{\perp}_{1}\rangle|^{2}
+\lambda^{-1}|\langle\psi|(\lambda A-iB)|\psi^{\perp}_{2}\rangle|^{2}:=\mathcal{L}_{1},
\end{align}
{\it which is valid for all $|\psi^{\perp}_{1}\rangle$ and $|\psi^{\perp}_{2}\rangle$ orthogonal to $|\psi\rangle$. If $-2i\langle[A, B]\rangle$ is negative then one changes the sign in Eq. (\ref{e:ineq}) to ensure the RHS is positive.}

The equality condition for Eq. (\ref{e:ineq})
holds if and only if $|\psi^{\perp}_{1}\rangle\propto(\widehat{A}+i\widehat{B})|\psi\rangle$ while $|\psi^{\perp}_{2}\rangle\propto(\lambda\widehat{A}+i\widehat{B})|\psi\rangle$. Denote the RHS of
Eq. (\ref{e:ineq}) by $\mathcal{L}_{1}$. Clearly $\mathcal{L}_{MP1}$ as a special case of $\mathcal{L}_{1}$,
as $lim_{\lambda\to 1}\mathcal{L}_{1}=\mathcal{L}_{MP1}$.
When $\lambda$ varies, one obtains a family of uncertainty relations and the lower
bounds $\mathcal{L}_{1}$ provide infinitely many uncertainty relations with weighted contributions for measurements $A$ and $B$. This will be advantageous when the ratio $\langle A\rangle/\langle B\rangle$ is not close to $1$.

See Methods for a proof of Theorem 1.

\vspace{2ex}
\leftline{\textbf{Theorem 2} {\it For arbitrary observables $A$, $B$ and any positive $\lambda$, we have the following weighted uncertainty relation:}}
\begin{align}\label{e:ineq2}
(1+\lambda)\Delta A^{2}+(1+\lambda^{-1})\Delta B^{2}
\geqslant|\langle\psi|A+B|\psi^{\perp}_{A+B}\rangle|^{2}+{\lambda}^{-1}|\langle\psi|(\lambda A-B)|\psi^{\perp}\rangle|^{2}:=\mathcal{L}_{2},
\end{align}
{\it where the equality holds if and only if  $|\psi^{\perp}\rangle\propto(\lambda \widehat{A}-\widehat{B})|\psi\rangle$}.

Denote the RHS of Eq. (\ref{e:ineq2}) by $\mathcal{L}_{2}$.  Note that the lower bound $\mathcal{L}_{2}$ is a nontrivial generalization of $\mathcal{L}_{MP2}$, as the latter
is a proper
bound unless $|\psi\rangle$ is an eigenstate of $A+B$. Even when $|\psi\rangle$ is an eigenstate of $A+B$, the new uncertainty bound $\mathcal{L}_{2}$ is also nonzero except for $\lambda=-1$ (Eq. (\ref{e:ineq2}) still holds for any nonzero real $\lambda$).
This means that in almost all cases the lower bound provided by
Eq. (\ref{e:ineq2}) is better except for $\lambda\neq-1$ and it compensates for the incompatibility
of the observables.
Obviously the bound $\mathcal{L}_{MP2}$
is a special case of $\mathcal{L}_{2}$ by canceling $|\langle\psi|(\lambda A-B)|\psi^{\perp}\rangle|^{2}$ when $\lambda=1$.

See Methods for a proof of Theorem 2.

Both lower bounds of the weighted uncertainty relations can be combined in a single uncertainty relation for the sum of variances:

\vspace{2ex}
\leftline{\textbf{Theorem 3} {\it For arbitrary observables $A$, $B$ and any positive number $\lambda$, we have the following weighted uncertainty relation:}}
\begin{equation}
(1+\lambda)\Delta A^{2}+(1+\lambda^{-1})\Delta B^{2}\geqslant\max(\mathcal{L}_{1}, \mathcal{L}_{2}).
\end{equation}

Theorems 1 and 2 provide a strengthened uncertainty relation and remove the limitation of the Maccone-Pati bounds. In fact, in the case when $|\psi\rangle$ is an eigenstate of $A$ or $B$, both Heisenberg-Robertson's and Schr\"{o}dinger's uncertainty relations are trivial, nevertheless our lower bound remains nonzero unless $|\psi\rangle$ is a common eigenstate of $A$ and $B$, but this is essentially equivalent to the classical
situation. It is also easy to see that if $|\psi\rangle$ is an eigenstate of $A\pm iB$,
$|\langle\psi|A\pm iB|\psi^{\perp}\rangle|^{2}$ in $\mathcal{L}_{MP1}$ will vanish while the term $|\langle\psi|(A-iB)|\psi^{\perp}_{1}\rangle|^{2}+\lambda^{-1}|\langle\psi|(\lambda A-iB)|\psi^{\perp}_{2}\rangle|^{2}$ in $\mathcal{L}_{1}$ is still nonzero unless $\lambda=1$.  Moreover, $\mathcal{L}_{MP2}$ will become null when $|\psi\rangle$ is an eigenstate of $A+B$, but at the same time $\mathcal{L}_{2}$ is still nontrivial.

Besides having a nontrivial bound in almost all cases, our weighted uncertainty relations can also lead to a tighter
bound for the sum of variances.
We give an algorithm to extract the usual uncertainty relation when
one of Maccone-Pati's relations becomes trivial. Choose two $\lambda_i$:
$\lambda_1>1>\lambda_2>0$ and enter our uncertainty relations Eq. (\ref{e:ineq}). Denote
$b_i=(1+\lambda_i)\Delta A^2+(1+\lambda_i^{-1})\Delta B^2$, then we have for $k=1, 2$
\begin{align}\label{e:para3}
\Delta A^2+\Delta B^2=\frac1{\lambda_1-\lambda_2}(\frac{\lambda_1(1-\lambda_2)}{1+\lambda_1}b_1+
\frac{\lambda_2(\lambda_1-1)}{1+\lambda_2}b_2)
\geqslant \frac1{\lambda_1-\lambda_2}(\frac{1-\lambda_2}{1+\lambda_1^{-1}}\mathcal{L}_k(\lambda_1)+
\frac{\lambda_1-1}{1+\lambda_2^{-1}}\mathcal{L}_k(\lambda_2)),
\end{align}
which always provides a nontrivial lower bound
for the sum of variances even when the state is an eigenvector of $A+B$. This clearly shows that
the weighted uncertainty relations can help recover the uncertainties and
remove the restriction placed in Maccone-Pati's uncertainty relation. Furthermore, taking the limit of $\lambda_i\to 1$ one has that
for $k=1, 2$
\begin{align}\label{e:para4}
\Delta A^2+\Delta B^2\geqslant \frac12lim_{\lambda\to 1}\mathcal{L}_k(\lambda).
\end{align}
For simplicity we refer to the RHS of Eq. (\ref{e:para3}) or the derived bound in Eq. (\ref{e:para4}) as {\it our lower bound of the sum of variances}, which usually is a multiple of
our bound from the weighted sum (see FIG. 1). In FIG 1 one will see that our bound
$\frac12\mathcal{L}_2$ derived in Eq. (\ref{e:para4}) is always tighter than the Maccone-Pati bound $\mathcal{L}_{MP2}$. In Eq. ({\ref{e:better}) we will use another method to show that our bound is tighter than Maccone-Pati's bound.

\begin{figure}\label{f:xy}
\centering
\includegraphics[width=0.45\textwidth]{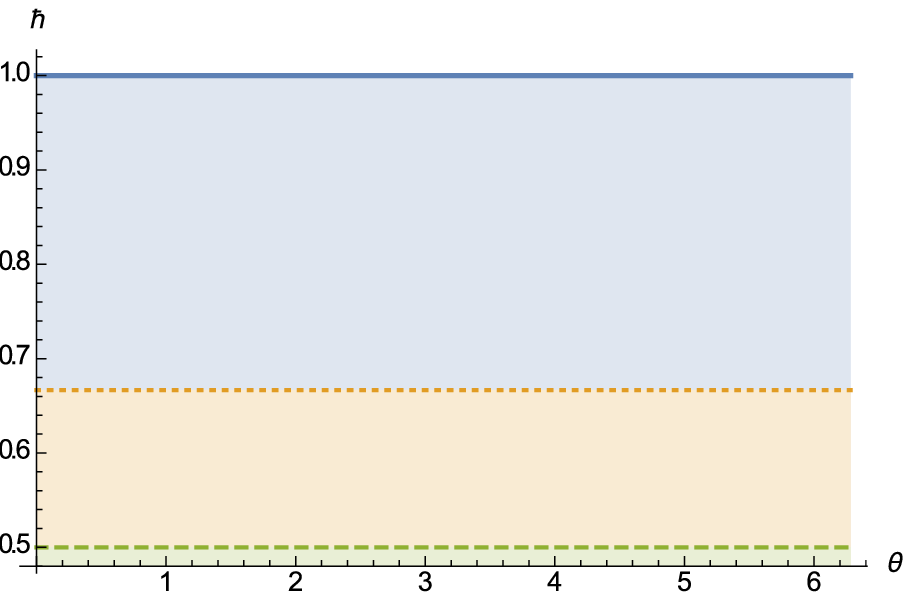}
\caption{Comparison of our bound $\frac12\mathcal{L}_2$ with Maccone-Pati's bound $\mathcal{L}_{MP2}$ for
operators $J_x$ and $J_y$ in a spin one system.
The top solid line is variance sum uncertainty $(\Delta J_{x})^{2}+(\Delta J_{y})^{2}$, the middle dotted line is $\frac{1}{2}\mathcal{L}_{2}$,
and the bottom dashed one is $\mathcal{L}_{MP2}$.}
\end{figure}

As an example to show our lower bound is tighter, we consider the spin one system with the
pure state
$|\psi\rangle=\cos \frac{\theta}{2}|0\rangle+\sin \frac{\theta}{2}|2\rangle$, $0\leqslant\theta<2\pi$.
Take the
angular momentum operators \cite{Rivas, Chen} with $\hbar=1$:
\begin{align}\label{e:angular}
J_{x}=\frac{1}{\sqrt{2}}
\left(
\begin{array}{ccc}
  0 & 1 & 0 \\
  1 & 0 & 1 \\
  0 & 1 & 0
\end{array}
\right), ~~~
J_{y}=\frac{1}{\sqrt{2}}
\left(
\begin{array}{ccc}
  0 & -i & 0 \\
  i & 0 & -i \\
  0 & i & 0
\end{array}
\right), ~~~
J_{z}=
\left(
\begin{array}{ccc}
  1 & 0 & 0 \\
  0 & 0 & 0 \\
  0 & 0 & -1
\end{array}
\right).
\end{align}
Direct calculation gives
$$(\Delta J_{x})^{2}=\frac{1}{2}(1+\sin\theta), ~~~ (\Delta J_{y})^{2}=\frac{1}{2}(1-\sin\theta), ~~~ (\Delta J_{z})^{2}=\sin^{2}\theta;$$
$$[\Delta(J_{x}+J_{y})]^{2}=1, [\Delta(J_{y}+J_{z})]^{2}=\frac{1}{2}(1-\sin\theta)+\sin^{2}\theta, [\Delta(J_{x}+J_{z})]^{2}=\frac{1}{2}(1+\sin\theta)+\sin^{2}\theta;$$
$$[\Delta(J_{x}+J_{y}+J_{z})]^{2}=1+\sin^{2}\theta.$$
To compare Macconne-Pati's uncertainty bound
$\mathcal{L}_{MP2}$ in Eq. (\ref{e:MP2}) with our bound $\frac{1}{2}\mathcal{L}_{2}$ in Eq. (\ref{e:ineq2}) (see also Eq. (\ref{e:para4})), setting $\lambda=1$ we get
$$(\Delta J_{x})^{2}+(\Delta J_{y})^{2}\geqslant\frac{1}{2}\mathcal{L}_{2}.$$
Also we have $(\Delta J_{x})^{2}+(\Delta J_{y})^{2}=1$ and $\mathcal{L}_{MP2}=\frac{1}{2}|\langle\psi^{\perp}_{A+B}|A+B|\psi\rangle|^{2}=\frac{1}{2}$. Suppose $|\psi^{\perp}\rangle=a|0\rangle+b|1\rangle+c|2\rangle$
with $|a|^{2}+|b|^{2}+|c|^{2}=1$. Using $\langle\psi|\psi^{\perp}\rangle=0$ we get  $$\frac{1}{2}\mathcal{L}_{2}=\frac{1}{2}+\frac{|b|^{2}}{2}\geqslant\mathcal{L}_{MP2}.$$
If we choose $a=\frac{1}{\sqrt{3}}, b=\frac{1}{\sqrt{3}}, c=-\frac{1}{\sqrt{3}}$, then $\frac{1}{2}\mathcal{L}_{2}=\frac{2}{3}$. Subsequently
$$(\Delta J_{x})^{2}+(\Delta J_{y})^{2}>\frac{1}{2}\mathcal{L}_{2}>\mathcal{L}_{MP2}.$$
On the other hand, if we set $a=0, b=1, c=0$ then $\frac{1}{2}\mathcal{L}_{2}=1=(\Delta J_{x})^{2}+(\Delta J_{y})^{2}>\mathcal{L}_{MP2}$. Clearly our bound $\frac{1}{2}\mathcal{L}_{2}$ is tighter than $\mathcal{L}_{MP2}$. The comparison is shown in FIG. 1.

We can also consider $(\Delta J_{y})^{2}+(\Delta J_{z})^{2}$, and direct computation shows $(\Delta J_{y})^{2}+(\Delta J_{z})^{2}=\frac{1}{2}+\sin^{2}\theta-\frac{1}{2}\sin\theta$, $\mathcal{L}_{MP2}=\frac{1}{2}(\frac{1}{2}+\sin^{2}\theta-\frac{1}{2}\sin\theta)$. Choose $|\psi^{\perp}\rangle=|1\rangle$ then $\frac{1}{2}\mathcal{L}_{2}=\frac{1}{2}-\frac{1}{2}\sin\theta+\frac{1}{2}\sin^{2}\theta$. Therefore
$$(\Delta J_{y})^{2}+(\Delta J_{z})^{2}\geqslant\frac{1}{2}\mathcal{L}_{2}\geqslant\mathcal{L}_{MP2}.$$
Apparently our bound $\frac{1}{2}\mathcal{L}_{2}$ is better than $\mathcal{L}_{MP2}$. FIG. 2. illustrates the comparison.
\begin{figure}\label{f:yz}
\centering
\includegraphics[width=0.45\textwidth]{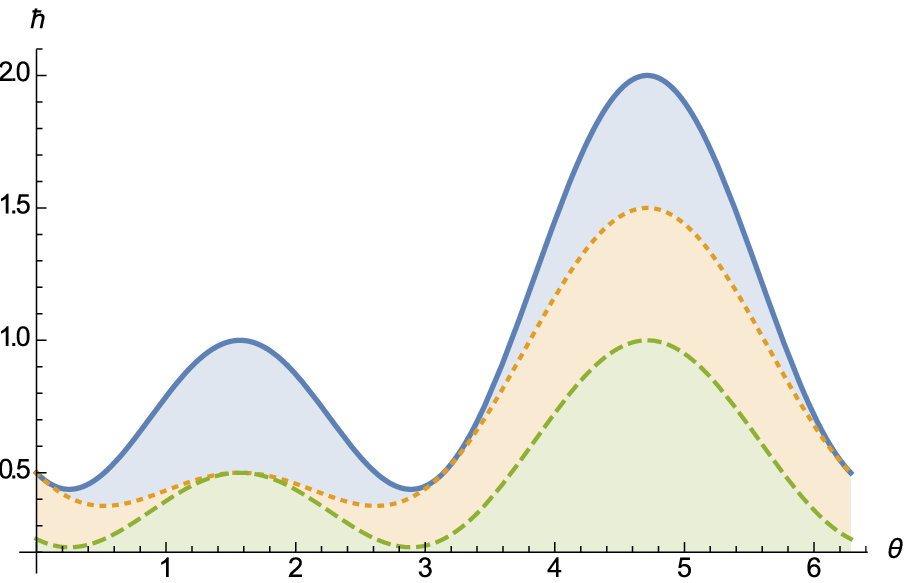}
\caption{Comparison of our bound $\frac{1}{2}\mathcal{L}_{2}$
with Maccone-Pati's bound $\mathcal{L}_{MP2}$ for operators $J_y$ and $J_z$ in a spin one system.
 The top solid curve is variance sum uncertainty $(\Delta J_{y})^{2}+(\Delta J_{z})^{2}$,
 the middle dotted curve is $\frac{1}{2}\mathcal{L}_{2}$ and the bottom dashed one is
 $\mathcal{L}_{MP2}$.}
\end{figure}

The bound $\mathcal{L}_{2}=\mathcal{L}_{2}(\lambda)$
is a function of $\lambda$. To analyze when $\mathcal{L}_{2}(\lambda)$
best approximates $(1+\lambda)\Delta A^{2}+(1+\lambda^{-1})\Delta B^{2}$, we define {\it the error function}
$f(\lambda)=(1+\lambda)\Delta A^{2}+(1+\lambda^{-1})\Delta B^{2}-\mathcal{L}_{2}(\lambda)$.
At an extremal point $\lambda_{0}$, the bound $\mathcal{L}_{2}(\lambda)$ is closest to the weighted sum
and one of the following two conditions must hold. Either $f'(\lambda_0)$ does not
exist or $f'(\lambda_{0})=\Delta A^{2}-\lambda^{-2}_{0}\Delta B^{2}-\mathcal{L}_{2}^{'}(\lambda_{0})=0$. If $\mathcal{L}_{2}^{'}(1)=\Delta A^{2}-\Delta B^{2}$, then $\lambda=1$ is the extremal point and we call it an {\it equilibrium point of the uncertainty relation}. In this case both observables $A$ and $B$ give
the same contribution to the uncertainty relation.
Usually $\lambda=1$ is not an extremal point, so in general observables $A$ and $B$ contribute unequally to the uncertainty relation.

To see an example of this phenomenon, let's consider again the quantum state
$|\psi\rangle=\cos \frac{\theta}{2}|0\rangle+\sin \frac{\theta}{2}|2\rangle\ \ (0<\theta<2\pi, \theta\neq\pi)$
and the angular momentum operators. Choose $|\psi^{\perp}\rangle=|1\rangle$, then
\begin{align}\label{e:f}
f(\lambda)=(1+\lambda)\Delta J_{y}^{2}+(1+\lambda^{-1})\Delta J_{z}^{2}-\mathcal{L}_{2}(\lambda)
=\lambda^{-1}\sin^{2}\theta,
\end{align}
while $f^{'}(\lambda)=-\lambda^{-2}\sin^{2}\theta<0$, hence $f(1)>f(\lambda), \forall \lambda>1$ (for fixed $\theta$). So for this $|\psi\rangle$,
$J_{y}$ and $J_{z}$ never contribute equally to the uncertainty relation,
which explains the need for a weighted uncertainty relation.
FIG. 3 shows the error function $f(\lambda)$ and $\mathcal{L}_{2}(\lambda)$.
In general $f$ is a function of both $\lambda$ and $\theta$, finding its extremal points involves a PDE equation.
For higher dimension quantum states or multi-operator cases, the situation is more complicated.
\begin{figure}
\centering
\includegraphics[width=0.45\textwidth]{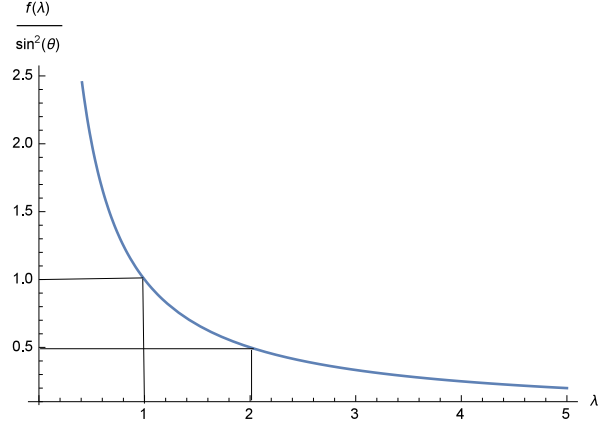}
\caption{Error function Eq. (\ref{e:f}) of Uncertainty Relation. The figure shows that the difference between uncertainty relation and its bound for fixed form $\mathcal{L}_{2}(\lambda)$ becomes less when $\lambda$ increases, which means that better estimation may be obtained through
larger $\lambda$.}
\end{figure}

In general, all variance-based sum uncertainty relations can mix in weights to provide an optimal lower bound. To compare the variance-based sum uncertainty relation with weighted uncertainty relation, take the lower bound $\mathcal{L}_{2}$ for a more detailed analysis: set $\lambda=1$ then $\Delta A^{2}+\Delta B^{2}
\geqslant\frac{1}{2}|\langle\psi|A+B|\psi^{\perp}_{A+B}\rangle|^{2}+\frac{1}{2}|\langle\psi|(A-B)|\psi^{\perp}\rangle|^{2}$, it is not only a typical variance-based sum uncertainty relation, but also provides a better lower bound than Maccone-Pati's lower bound $\mathcal{L}_{MP2}$. Moreover, this lower bound can be further improved by a
mixture of weights.

\vspace{2ex}
\leftline{\textbf{Corollary 1} {\it For arbitrary observables $A$, $B$ and any positive number $\lambda$, we have the following weighted uncertainty relation:}}
\begin{align}\label{e:better}
\Delta A^{2}+\Delta B^{2}
\geqslant&\sup\limits_{\lambda}[|\langle\psi|\frac{1}{\sqrt{1+\lambda}}A+\frac{1}{\sqrt{1+\lambda^{-1}}}B|
\psi^{\perp}_{\frac{A}{\sqrt{1+\lambda}}+\frac{B}{\sqrt{1+\lambda^{-1}}}}\rangle|^{2}
+{\lambda}^{-1}|\langle\psi|(\frac{\lambda}{\sqrt{1+\lambda}}A-\frac{1}{\sqrt{1+\lambda^{-1}}}B)|\psi^{\perp}\rangle|^{2}]\notag\\
\geqslant&\frac{1}{2}|\langle\psi|A+B|\psi^{\perp}_{A+B}\rangle|^{2}+\frac{1}{2}|\langle\psi|(A-B)|\psi^{\perp}\rangle|^{2},
\end{align}
{\it where $|\psi^{\perp}_{\frac{A}{\sqrt{1+\lambda}}+\frac{B}{\sqrt{1+\lambda^{-1}}}}\rangle\propto
(\frac{A}{\sqrt{1+\lambda}}+\frac{B}{\sqrt{1+\lambda^{-1}}}-\langle \frac{A}{\sqrt{1+\lambda}}+\frac{B}{\sqrt{1+\lambda^{-1}}} \rangle)|\psi\rangle)$ is a state orthogonal to $|\psi\rangle$.}

Through Eq. (\ref{e:better}), it is easy to see $\frac{1}{2}|\langle\psi|A+B|\psi^{\perp}_{A+B}\rangle|^{2}+\frac{1}{2}|\langle\psi|(A-B)|\psi^{\perp}\rangle|^{2}$ is the special case of $\lambda=1$ and, {\it a fortiori}, the lower bound with weights is tighter than the standard one.

See Methods for a proof of Corollary 1.

One can study the general weighted sum of variances $x\Delta A^{2}+y\Delta B^{2}$
based on the special weighted sum $(1+\lambda)\Delta A^{2}+(1+\lambda^{-1})\Delta B^{2}$.
Theorem 4 details the relationship between the general and special weighted sum uncertainty relations.

\vspace{2ex}
\leftline{\textbf{Theorem 4} {\it For arbitrary observables $A$, $B$ and $x$, $y$ such that
$xy(x+y)>0$, the following weighted uncertainty relation holds.}}
\begin{align}\label{e:arbitrary}
x\Delta A^{2}+y\Delta B^{2}\geqslant\frac{xy}{x+y}\mathcal{L}_{2}(\frac{x}{y}).
\end{align}

See Methods for a proof of Theorem 4.

According to Deutsch \cite{Deutsch}, uncertainty in the result of a measurement of observables $A$ and $B$ should be quantified as
an inequality with certain lower bound. One can seek such a bound in a general form $\mathcal{U}(A, B, |\psi\rangle)$ which may
not simply be a sum or product by weighted uncertainty relations.
For instance, we take $\mathcal{U}(A, B, |\psi\rangle)=\frac{1}{1-\Delta A}+e^{\Delta B}$, its bound can be extracted from
Theorem 4.

\vspace{2ex}
\leftline{\textbf{Remark 1}
{\it For $\mid\Delta A\mid<1$ and arbitrary observable $B$, $\frac{1}{1-\Delta A}+\exp(\Delta B)$ has a nonnegative lower bound:}}
\begin{align}\label{e:example}
\frac{1}{1-\Delta A}+e^{\Delta B}\geqslant\sum\limits_{n=0}^{\infty}[\frac{\mathcal{L}_{2}(\sqrt[n]{n!})}{2(\sqrt[n]{n!}+1)}]^{n}.
\end{align}

See Methods for a proof of Remark 1.

We now generalize the weighted uncertainty relations to multi-operator cases. To emphasize our point, we
recall the trivial generalization from Maccone-Pati's lower bound.

\vspace{2ex}
\leftline{\textbf{Lemma 1} {\it For arbitrary observables $A_{i}$ ($i=1, \ldots, n$), we have the following variance-based sum uncertainty relation:}}
\begin{align}\label{e:ineqs}
\sum\limits_{i}\Delta A_{i}^{2}\geqslant\frac{1}{n}\Delta S^{2}
=\frac{1}{n}|\langle\psi|S|\psi^{\perp}_{S}\rangle|^{2},
\end{align}
{\it where $|\psi^{\perp}_{S}\rangle\propto(S-\langle S\rangle)|\psi\rangle$ is a unit state
perpendicular to $|\psi\rangle$ while $S=\sum\limits_{i}A_{i}$. The RHS of Eq. (\ref{e:ineqs}) is nonzero unless $|\psi\rangle$ is an eigenstate of $S=\sum\limits_{i}A_{i}$.}

See Methods for a proof of Lemma 1.

Notice that $|\psi\rangle$ can be an eigenstate of $\sum\limits_{i}A_{i}$ without being
that of any $A_{i}$, in which case the lower bound is still trivial. However, the bound is not optimal
and sometimes becomes trivial when the observables
are incompatible in the general situation. We now introduce {\it generalized weighted uncertainty relations} to deal with these drawbacks.

\vspace{2ex}
\leftline{\textbf{Theorem 5} {\it For arbitrary $n$ observables $A_{i}$ and positive numbers $\lambda_{i}$, we have following sum uncertainty relation:}}
\begin{align}\label{e:mineq0}
\sum\limits_{i, j=1}^{n}\frac{\lambda_{i}}{\lambda_{j}}&\Delta A_{i}^{2}
\geqslant|\langle\psi|S|\psi^{\perp}_{0}\rangle|^{2}+\sum\limits_{1\leqslant i<j\leqslant n}|\langle\psi|(\sqrt{\frac{\lambda_{i}}{\lambda_{j}}}A_{i}-\sqrt{\frac{\lambda_{j}}{\lambda_{i}}}A_{j})|\psi^{\perp}_{ij}\rangle|^{2},
\end{align}
{\it where $|\psi^{\perp}_{ij}\rangle\propto(\sqrt{\frac{\lambda_{i}}{\lambda_{j}}}\widehat{A}_{i}
-\sqrt{\frac{\lambda_{j}}{\lambda_{i}}}\widehat{A}_{j})|\psi\rangle$ and $|\psi^{\perp}_{0}\rangle$ is
any unit state $\perp|\psi\rangle$.}

See Methods for a proof of Theorem 5.

The RHS $\mathcal{L}_{0}$ of (\ref{e:mineq0}) depends on the choice of $\lambda_{i}$. By the same trick and fixing the $(i, j)$-term of Eq. (\ref{e:mineq0}), we arrive at

\vspace{2ex}
\leftline{\textbf{Theorem 6} {\it For arbitrary $n$ observables $A_{i}$ and positive numbers $\lambda_{i}$, we have following sum uncertainty relation:}}
\begin{align}\label{e:mineqij}
\sum\limits_{i, j=1}^{n}\frac{\lambda_{i}}{\lambda_{j}}&\Delta A_{i}^{2}
\geqslant|\langle\psi|S|\psi^{\perp}_{S}\rangle|^{2}+
|\langle\psi|(\sqrt{\frac{\lambda_{i}}{\lambda_{j}}}A_{i}-\sqrt{\frac{\lambda_{j}}{\lambda_{i}}}A_{j})|\psi^{\perp}_{ij}\rangle|^{2}
+\sum\limits_{1\leqslant k\neq i<l\neq j\leqslant n}|\langle\psi|(\sqrt{\frac{\lambda_{k}}{\lambda_{l}}}A_{k}-\sqrt{\frac{\lambda_{l}}{\lambda_{k}}}A_{l})|\psi^{\perp}_{kl}\rangle|^{2},
\end{align}
{\it where $|\psi^{\perp}_{ij}\rangle$ is orthogonal to $|\psi\rangle$,
$|\psi^{\perp}_{S}\rangle\propto(S-\langle S\rangle)|\psi\rangle$, and $|\psi^{\perp}_{kl}\rangle\propto(\sqrt{\frac{\lambda_{k}}{\lambda_{l}}}\widehat{A}_{k}
-\sqrt{\frac{\lambda_{l}}{\lambda_{k}}}\widehat{A}_{l})|\psi\rangle$.}

Clearly, $\mathcal{L}_{0}$ and all the RHS $\mathcal{L}_{ij}$ of Eq. (\ref{e:mineqij})
comes form Theorem 5. and Theorem 6. respectively
can be combined into a single uncertainty relation for variances:

\vspace{2ex}
\leftline{\textbf{Theorem 7} {\it For arbitrary $n$ observables $A_{i}$ and any positive numbers $\lambda_{i}$, we have the following sum uncertainty relation:}}
\begin{equation}\label{e:bound}
\sum\limits_{i, j=1}^{n}\frac{\lambda_{i}}{\lambda_{j}}\Delta A_{i}^{2}\geqslant\max\limits_{1\leqslant i<j\leqslant n}(\mathcal{L}_{ij}, \mathcal{L}_{0}).
\end{equation}

When setting $\lambda_{i}=\lambda_{j}$, the RHS of Eq. (\ref{e:bound}) is still stronger than Eq. (\ref{e:ineqs}), since it keeps all the terms $\sum\limits_{1\leqslant i<j\leqslant n}\parallel(\sqrt{\frac{\lambda_{i}}{\lambda_{j}}}\widehat{A}_{i}-\sqrt{\frac{\lambda_{j}}{\lambda_{i}}}\widehat{A}_{j})|\psi\rangle\parallel^{2}$ appearing in Eq. (\ref{e:mineq0}). We remark that a default choice
of $|\psi^{\perp}\rangle$ in Eq. (\ref{e:mineqij}) is by Vaidman's formula \cite{Vaidman, Goldenberg}: $|\psi^{\perp}_{S}\rangle=(S-\langle S\rangle)|\psi\rangle/\Delta S$. We can select suitable $\lambda_{i}$ such that $\max(\mathcal{L}_{ij}, \mathcal{L}_{0})$ is nontrivial. They are zero if and only if $|\psi\rangle$ is a common eigenstate of all observables, which happens only when the system is equivalent to the classical situation.
In this sense our weighted uncertainty relation can handle
all possible quantum situations.

If two or more terms in the RHS of these equality are replaced by the Cauchy-Schwarz's inequality
simultaneously, the corresponding lower bound can not be bigger than the one by replacing just one term.
In other words, $\max(\mathcal{L}_{ij}, \mathcal{L}_{0})$ is better than the lower bounds by changing
more than one term. The LHS of Eq. (\ref{e:bound}) has only positive coefficients since $\lambda_{i}$ are positive.

\section*{Discussions}

There are several physical motivations and mathematical considerations behind our method. 
First, to remove the restriction of one of Macconne-Pati's uncertainty relations
(i.e. when $\psi$ is
an eigenstate of $A+B$) and recover the lower bound for $\Delta^2(A)+\Delta^2(B)$, we
consider a perturbation of $A$ and $B$, or rather, $A'=\sqrt{1+\lambda}A$, $B'=\sqrt{1+\lambda^{-1}}B$ ($\lambda>0$).
Then
\begin{equation}\label{rescaled}
\Delta^2(A')+\Delta^2(B')=(1+\lambda)\Delta^2(A)+(1+\lambda^{-1})\Delta^2(B).
\end{equation}
This means that the lower bound of the sum of variances can be obtained by scaled observables.
Actually with the given measurement data of the variances, it is easy to compute the lower bound
using our new formula. This is in line with the general strategy of perturbation method,
just as many singular properties can be better studied through deformation.

Secondly, the idea of the weighted sum or average is similar to well-known techniques used in
both statistical mechanics and mathematical physics. Through the weighted averages one
may know better about the whole picture in an unbiased way.

Thirdly, the weighted sum is actually a $q$-deformation of the original sum of variances. In fact,
the sum $2\Delta^2(A)+2\Delta^2(B)$ is deformed to
$$[2]\lambda^{1/2}\Delta^2(A)+[2]\lambda^{-1/2}\Delta^2(B),$$
where $[2]=\lambda^{1/2}+\lambda^{-1/2}$ is the quantum integer of $2$ used widely in quantum groups, Yang-Baxter equations,
and quantum integrable systems or statistical mechanics. The opposite phase factors $\lambda^{\pm 1/2}$ in front of the variances reflect a balance of the weighted distribution.

Last but not the least, the usual sum of variances can be solved from our weighted sums (see Eqs. (\ref{e:para3}-\ref{e:para4})),
and the derived bound is proved to be tighter than the original Maccone-Pati's bound. Noted that, there is another result connect with
uncertainty relations in terms of weighted sums of variances, for more detail, see \cite{Lars}.

\section*{Conclusions}

The Heisenberg-Robertson and Schr\"{o}dinger uncertainty relations have been skillfully 
generalized
by Maccone and Pati in order to capture the concept of incompatibility
of the observables $A$ and $B$ on the quantum system $|\psi\rangle$. Although
other generalizations of Maccone-Pati's relations have been considered \cite{Sun} by
refining the RHS, our generalization provides a non-trivial lower bound in all
quantum situations. One of Maccone-Pati's
relations becomes trivial when $|\psi\rangle$ is an eigenstate of $A+B$. To remove the
restriction of their relation,
we have proposed
a weighted uncertainty relation to obtain a better lower bound for the sum of the variances.
The parametric uncertainty relations form a family of Bohr-type inequalities and take into account of
individual contribution from the observables so that they are nontrivial in almost all
cases except when $|\psi\rangle$ is a common eigenstate of all observables. In particular, Maccone-Pati's uncertainty relations are special cases of our deformed weighted uncertainty
relations. Furthermore, we have shown that the sum of variances
can be extracted from our weighted sums and our derived bound is always tighter than Maccone-Pati bound $\mathcal{L}_{MP2}$
(see discussion before Eq. (\ref{e:para4})).
 We have also derived weighted uncertainty relations for multi-observables
and the lower bound has been proved to be optimal in all quantum cases.

\section*{Methods}

\textbf{Proof of Theorem 1}
We start by recalling the parallelogram law in Hilbert space.
Let $A$ and $B$ be two observables and $|\psi\rangle$ a fixed quantum state. One has that
\begin{equation}
2\Delta A^{2}+2\Delta B^{2}=\parallel(\widehat{A}+\alpha\widehat{B})|\psi\rangle\parallel^{2}+\parallel(\widehat{A}-\alpha\widehat{B})|\psi\rangle\parallel^{2},
\end{equation}
for any $|\alpha|=1$. Since $\Delta(A+B)=\parallel(\widehat{A}+\widehat{B})|\psi\rangle\parallel$, $\Delta(A-B)=\parallel(\widehat{A}-\widehat{B})|\psi\rangle\parallel$, we can obtain Eq. (\ref{e:MP1}) when $\alpha=\pm i$ and Eq. (\ref{e:MP2}) when $\alpha=1$. Note that $\mathcal{L}_{MP2}$ may be zero even if
$A$ and $B$ are incompatible. For example this happens if $|\psi\rangle$ is an eigenstate of $A+B$.
Our idea is to consider a perturbation of $A+B$, or $A$ and $B$ to fix this. We consider the generalized parallelogram law in Hilbert space in the following form:
\begin{align}\label{e:para1}
(1+\lambda)\Delta A^{2}+(1+\lambda^{-1})\Delta B^{2}
=\parallel(\widehat{A}-\alpha\widehat{B})|\psi\rangle\parallel^{2}&+{\lambda}^{-1}\parallel(\lambda \widehat{A}+\alpha\widehat{B})|\psi\rangle\parallel^{2},
\end{align}
where $\lambda$ is a nonzero real number and $\alpha\in\mathbb C$ with modulus one.
In fact, the identity can be easily verified by expanding
$\Delta(A-\alpha B)^2$ and $\lambda^{-1}\Delta(\lambda A+\alpha B)^2$ using
$\Delta(A)^2=\langle\psi|\widehat{A}^2|\psi\rangle$.

We now derive the {\it weighted uncertainty relation} in the form
$(1+\lambda)\Delta A^{2}+(1+\lambda^{-1})\Delta B^{2}$. Since
$\parallel(\widehat{A}-i\widehat{B})|\psi\rangle\parallel^{2}=-2i\langle[A, B]\rangle+\parallel(\widehat{A}+i\widehat{B})|\psi\rangle\parallel^{2}$, combine with Cauchy-Schwarz inequality completes the proof.
\vspace{2ex}

\leftline{\textbf{Proof of Theorem 2} If we set $\alpha=-1$ in Eq. (\ref{e:para1}), then we get the result directly.}
\vspace{2ex}

\leftline{\textbf{Proof of Corollary 1} For $\lambda>0$, set $A^{'}=\sqrt{1+\lambda}A, B^{'}=\sqrt{1+\lambda^{-1}}B$ (see Eq. (\ref{rescaled})), so}
\begin{align}\label{e:change}
\mathcal{L}_{2}(\lambda)=|\langle\psi|\frac{1}{\sqrt{1+\lambda}}A^{'}+\frac{1}{\sqrt{1+\lambda^{-1}}}B^{'}|\psi^{\perp}_{A+B}\rangle|^{2}
+{\lambda}^{-1}|\langle\psi|(\frac{\lambda}{\sqrt{1+\lambda}}A^{'}-\frac{1}{\sqrt{1+\lambda^{-1}}}B^{'})|\psi^{\perp}\rangle|^{2},
\end{align}
where the RHS $\mathcal{L}_{2}(\lambda, A^{'}, B^{'})$ satisfies that $\sup_{\lambda}\mathcal{L}_{2}(\lambda, A^{'}, B^{'})\geqslant \mathcal{L}_{2}(1, A^{'}, B^{'})$ which implies that the weighted uncertainty relation is better than the ordinary sum:
$(\Delta A^{'})^{2}+(\Delta B^{'})^{2}\geqslant\sup_{\lambda}\mathcal{L}_{2}(\lambda, A^{'}, B^{'})\geqslant \mathcal{L}_{2}(1, A^{'}, B^{'})$. Followed by parameter transformation, we get Eq. (\ref{e:better}).
\vspace{2ex}

\leftline{\textbf{Proof of Theorem 4} For arbitrary weighted uncertainty relation $x\Delta A^{2}+y\Delta B^{2}$, denote $f(x, y)= \frac{x+y}{xy}>0$, then}
\begin{align}
x\Delta A^{2}+y\Delta B^{2}=\frac{1}{f(x, y)}[f(x, y)x\Delta A^{2}+f(x, y)y\Delta B^{2}].
\end{align}
Set $\lambda=f(x, y)x-1=\frac{x}{y}$, then $\lambda^{-1}=f(x, y)y-1$. Thus
\begin{align}
x\Delta A^{2}+y\Delta B^{2}=\frac{1}{f(x, y)}[(1+\lambda)\Delta A^{2}+(1+\lambda^{-1})\Delta B^{2}]
\geqslant\frac{1}{f(x, y)}\mathcal{L}_{2}(x/y).
\end{align}
\vspace{2ex}

\leftline{\textbf{Proof of Remark 1} Since}
\begin{align}
\frac{1}{1-\Delta A}+e^{\Delta B}=\sum\limits_{n=0}^{\infty}[(\Delta A)^{n}+(\frac{1}{\sqrt[n]{n!}}\Delta B)^{n}]
\geqslant\sum\limits_{n=0}^{\infty}\frac{1}{2^n}(\Delta A+\frac{1}{\sqrt[n]{n!}}\Delta B)^{n},
\end{align}
with $x=1$, $y=\frac{1}{\sqrt[n]{n!}}$, $\lambda=\sqrt[n]{n!}$ and $f(x, y)=\frac{x+y}{xy}$, we get
\begin{align}
\Delta A+\frac{1}{\sqrt[n]{n!}}\Delta B=\frac{1}{f(x, y)}[(1+\lambda)\Delta A+(1+\lambda^{-1})\Delta B]
\geqslant(\sqrt[n]{n!}+1)^{-1}\mathcal{L}_{2}(\sqrt[n]{n!}),
\end{align}
thus
\begin{align}
\frac{1}{1-\Delta A}+e^{\Delta B}\geqslant\sum\limits_{n=0}^{\infty}[\frac{\mathcal{L}_{2}(\sqrt[n]{n!})}{2(\sqrt[n]{n!}+1)}]^{n}.
\end{align}
The right-hand is a positive lower bound of uncertainty relation $\frac{1}{1-\Delta A}+\exp(\Delta B)$.
\vspace{2ex}

\leftline{\textbf{Proof of Lemma 1} We recall Maccone-Pati's lower bound $\mathcal{L}_{MP2}$ using a different method.}
Note that
 $2\Delta A\Delta B\leqslant\Delta A^{2}+\Delta B^{2}$ and $\Delta(A+B)\leqslant\Delta A+\Delta B$,
 therefore $\Delta A^{2}+\Delta B^{2}\geqslant\frac{1}{2}\Delta(A+B)^{2}$. The physical meaning
  is that the total ignorance of an ensemble of quantum states
  is less than or equal to
 the sum of individual ignorance. This means that the sum of uncertainties
 obeys the convexity property \cite{Sahu}:
\begin{equation}\label{convex}
\Delta(\sum\limits_{i=1}^nA_{i})\leqslant\sum\limits_{i=1}^n\Delta A_{i}.
\end{equation}
Let $S=\sum\limits_{i}A_{i}$. It follows from Eq. (\ref{convex}) that
\begin{align}
\sum\limits_{i}\Delta A_{i}^{2}&\geqslant\frac{1}{n}\Delta S^{2}
=\frac{1}{n}|\langle\psi|S|\psi^{\perp}_{S}\rangle|^{2},
\end{align}
where $|\psi^{\perp}_{S}\rangle\propto(S-\langle S\rangle)|\psi\rangle$ is a unit state
perpendicular to $|\psi\rangle$.
\vspace{2ex}

\leftline{\textbf{Proof of Theorem 5} Using the generalized parallelogram law and Bohr's inequality \cite{Egecioglu, Hirzallah, Kato, Zeng, Mohammad, Zhang}, we obtain the following relation:}
\begin{align}\label{e:mineq}
\sum\limits_{i, j=1}^{n}\frac{\lambda_{i}}{\lambda_{j}}\Delta A_{i}^{2}=\parallel\widehat{S}|\psi\rangle\parallel^{2}
+\sum\limits_{1\leqslant i<j\leqslant n}\parallel(\sqrt{\frac{\lambda_{i}}{\lambda_{j}}}\widehat{A}_{i}-\sqrt{\frac{\lambda_{j}}{\lambda_{i}}}\widehat{A}_{j})|\psi\rangle\parallel^{2},
\end{align}
where $S=\sum\limits_{i}^{n}A_{i}$, $\widehat{S}=S-\langle S\rangle$ and $\lambda_{1}, \ldots, \lambda_{n}$ are positive real numbers. Combining with Cauchy-Schwarz inequality, we derive Eq. (\ref{e:mineq0}).

\section*{Acknowledgements}

We are grateful to Jian Wang and Yinshan Chang for
fruitful discussions. We also thank Chunqin Zhou
and Huihui Qin for their interest in the work.
The project is supported by
NSFC (grant Nrs. 11271138, 11531004), CSC and Simons Foundation 198129.

\section*{Author contributions statement}

Y. X. and N. J. wrote the main manuscript text. They also analysed the results and reviewed the manuscript
together with X. L.-J. and S.-M. F.

\section*{Additional information}

\textbf{Competing financial interests:} The authors declare no competing financial interests.
\vspace{1ex}

\leftline{\textbf{How to cite this article:} Xiao, Y., Jing, N., Li-Jost, X. and Fei, S.-M. Weighted Uncertainty Relations.}

\end{document}